# A compact and cost-effective laser-powered speckle visibility spectroscopy (SVS) device for measuring cerebral blood flow

Yu Xi Huang,[a,†] Simon Mahler,[a,†,*] Maya Dickson,[a] Aidin Abedi,[b] Julian M. Tyszka,[c] Yu Tung Lo,[b] Jonathan Russin,[b] Charles Liu,[b,*] Changhuei Yang[a,**]

[a] Department of Electrical Engineering, California Institute of Technology, Pasadena, California 91125, USA
[b] USC Neurorestoration Center and the Departments of Neurosurgery and Neurology, University of Southern California; Los Angeles, CA 90033, USA
[c] Division of Humanities and Social Sciences, California Institute of Technology, Pasadena, California 91125, USA
[d] Rancho Los Amigos National Rehabilitation Center, Downey CA, 90242, USA
[†]These authors contributed equally to this work.
[*]**Email:** cliu@usc.edu
[**]**Email:** chyang@caltech.edu

**Abstract**

In the realm of cerebrovascular monitoring, primary metrics typically include blood pressure, which influences cerebral blood flow (CBF) and is contingent upon vessel radius. Measuring CBF non-invasively poses a persistent challenge, primarily attributed to the difficulty of accessing and obtaining signal from the brain. This study aims to introduce a compact speckle visibility spectroscopy (SVS) device designed for non-invasive CBF measurements, offering cost-effectiveness and scalability while tracking CBF with remarkable sensitivity and temporal resolution. The wearable hardware has a modular design approach consisting solely of a laser diode as the source and a meticulously selected board camera as the detector. They both can be easily placed on a subject's head to measure CBF with no additional optical elements. The SVS device can achieve a sampling rate of 80 Hz with minimal susceptibility to external disturbances. The device also achieves better SNR compared with traditional fiber-based SVS devices, capturing about 70 times more signal and showing superior stability and reproducibility. It is designed to be paired and distributed in multiple configurations around the head, and measure signals that exceed the quality of prior optical CBF measurement techniques. Given its cost-effectiveness, scalability, and simplicity, this laser-centric tool offers significant potential in advancing non-invasive cerebral monitoring technologies.

## 1 Introduction

The brain stands as the most complicated and indispensable organ within the human body. Monitoring the cerebral blood flow (CBF) non-invasively bears significance in both clinical settings and cognitive neuroscience research[1]. Measuring CBF non-invasively poses a persistent challenge, primarily attributed to the difficulty of accessing and obtaining signal from the brain, especially in biomedical context where the exposure levels are restricted for the safety of the subjects[2]. As a result, efforts have been devoted in diverse methods for measuring CBF. Some notable techniques include transcranial Doppler ultrasound[3,4], magnetic resonance imaging (MRI)[5,6], near-infrared spectroscopy[7,8], electroencephalography[9], or cerebral oximetry. Optical monitoring of CBF stands to be expected more sensitive than other techniques[10], as it can better penetrate through skulls and tissues while providing higher temporal resolution.

Diffusing wave spectroscopy utilizing laser light transmitted through a scattering medium to extract the dynamic information has recently garnered attention as a promising tool for CBF monitoring[11–17]. One advantage of diffusing wave spectroscopy is the capability to collect a substantial number of photons that have interacted with the brain. It also presents numerous operational benefits, including its non-ionizing, safe radiation, straightforward methodology, use of relatively lightweight and cost-effective equipment, and compatibility with advanced commercial optical systems that can be readily adapted. In diffusing wave spectroscopy scheme, laser light is injected into the head using a laser source, and the emerging light is collected by a detector positioned at a source-to-detector (S-D) separation distance from the injection spot. The movements of blood cells within the travelling light's path will scatter and change the effective optical path lengths, resulting in a fluctuating laser speckle field.

There exist two types of sampling techniques to infer the blood flow: temporal and spatial. The temporal sampling technique, called time-domain diffuse correlation spectroscopy, is based on the use of the temporal ensemble of the speckle field and uses a photodetector working at a high frame rate (typically above 100 kHz) on a single (or on a small group of) speckle[11,12]. The spatial sampling technique is an off-shoot of laser speckle contrast imaging (LSCI) and is based on the use of spatial ensemble of the speckle field, usually referred as speckle visibility spectroscopy (SVS)[15,16,18–20] or as speckle contrast optical spectroscopy (SCOS)[17,21,22].

In SVS, instead of a high frame rate detecting device, a camera with a larger detecting area and a large number of pixels is used to collect more photons within the same frame[15,16,18]. The camera is typically working at an exposure time longer than the decorrelation time of the speckle field. This results in multiple different speckle patterns summing up onto a single camera frame. As the speckle field fluctuates, the speckle pattern recorded by the camera is smeared and washed out within the exposure time. Because the smearing or the washing out effect is due to the dynamics of the blood cells, the decorrelation time can be calculated from the degree of blurring of the captured frame, typically by calculating the speckle contrast. SVS was applied on the human head to monitor cerebral blood flow non-invasively, allowing for the detection of a larger number of speckles and an increased proportion of detected light from the brain[15,16].

This paper reports a compact SVS device designed for monitoring CBF. This wearable hardware consists solely of a continuous-wave laser diode and a high-resolution CMOS-based board camera that can be easily placed on a subject's head to measure CBF with no external optical elements. It offers real-time CBF monitoring at 80 Hz sampling rate while maintaining a lightweight and budget-friendly design. While similar wearable optical system was recently used to measure the changes of CBF during breath hold maneuver[23], this paper presents the design and processing of compact SVS and compares its gain in stability and SNR over fiber-based SVS.

We expect the compact SVS device to have certain advantages over fiber-based SVS devices. First, we show that compact SVS achieves better SNR compared with fiber-based SVS devices by collecting a larger amount of photons due to a significant increase in the detecting area and numerical aperture. Specifically, we measured the compact SVS version to capture about 70 times more signal relative to a comparable fiber-based SVS device, improving detectability of CBF at extended S-D distances. Second, it eliminates the motional artifacts associated with the light guide running from the head to the camera, showing a superior stability and reproducibility. Typically, when a large-diameter multimode fibers is used to collect the photons from the head to the camera, slight movements of the fiber can cause significant speckle changes, disrupting the results[17,19].

This issue is currently mitigated by minimizing fiber perturbations with extraordinary measures, which is not ideal.

The paper is organized as follows. First, we detail the design and experimental arrangement of the proposed compact SVS system and describe the data processing for calculating blood flow from the recorded camera images. Second, we compare the performance of compact SVS and fiber SVS by using static and moving phantoms. Finally, we experimentally compare the CBF measured from compact SVS and fiber SVS devices at different S-D distances from a cohort of five subjects. Our results show significant improvements in CBF measurement with the compact SVS over the fiber SVS device.

## 2  Methods

The arrangement of our compact SVS device is shown in Fig. 1. The system design is shown in Fig. 1(a) with the schematics shown on the left and a photograph of the 3D printed device shown on the right. The system includes a laser source for illumination and a board camera for detection. A dime is included in the photograph for size comparison.

For this study, we used a single-mode continuous wave 785 nm laser diode as a source [Thorlabs L785H1] which can deliver up to 200 mW. To ensure control over the illumination spot size and prevent undesirable laser light reflections or stray light, we housed the laser diode within a 3D-printed mount. The mounts were printed using black resin which absorbs light, minimizing back reflection and stray light. The laser diode was set several millimeters away from the skin of participants such that the illumination spot diameter was 5 mm[16]. The total illumination power was limited to 45 mW to ensure that the laser light intensity level of the area of illumination is well within the American National Standards Institute (ANSI) laser safety standards for maximum permissible exposure (2.95 mW/mm$^2$) for skin exposure to a 785nm laser beam[2].

At a specific S-D distance from the illumination spot, the detector was positioned on the head of the subject to collect the emerging light away from the laser illumination spot, Figs. 1(b) and 1(c). The collected laser light was directed onto a carefully selected camera equipped with a large sensor area and small pixel size, maximizing the number of speckles captured. We used a USB-board camera [Basler daA1920-160um (Sony IMX392 sensor)] as the detector. For optimal performance and stability, we typically operated the camera at a framerate of 80 frame-per-second (fps). The compact SVS system has the potential to achieve a sampling rate of up to 160 fps. However, it is capped at 80 fps to provide a balance between storage space and temporal resolution. To ensure time-synchronization among all pixels, the camera was configured with a global shutter setting. This camera features a pixel pitch of 3.4 µm, which offers a balance between the average intensity per pixel and the number of speckles per pixel which was measured to be about 10 speckles per pixels.

The depth to which the photons have travelled deep into the head is related to the S-D distance[13,16]. By tuning the S-D distance, one can tune the depth of penetration into the head, where a banana-shaped spatial sensitivity of the light path is usually observed as shown in Fig. 1(b)[13,16]. As the S-D distance increases, the banana-shape extends deeper into the brain, with deeper brain regions being more challenging to access. The spatial distribution of the exiting photons collected by a camera exhibit a granular pattern with areas of high and low intensity called speckles. The motions within the light paths, primarily due to the movement of red blood cells, will scatter and change the effective optical path lengths resulting in a fluctuating speckle field that varies in time.

Speckles arise from interference between the numerous random scatterings with the coherent light field and constitute a vast area of research[24]. We image these speckles onto a camera with a finite exposure time, Fig. 2(a). The camera must operate at a high enough frame rate to temporally resolve the dynamics, typically above 20 fps for blood flow measurements. Speckles undergo dynamic changes with a specific temporal evolution[25–27], characterized by the decorrelation time $\tau_c$ of the speckle field[28,29]. Typically, the camera is configured with an exposure time $T$ that is significantly larger than the decorrelation time $\tau_c$. As the speckle field fluctuates, the recorded speckled image would be smeared and washed out: the shorter the speckle decorrelation time, the more washed out the image. The dynamics of the speckles can be quantified by calculating the speckle contrast of the recorded image.

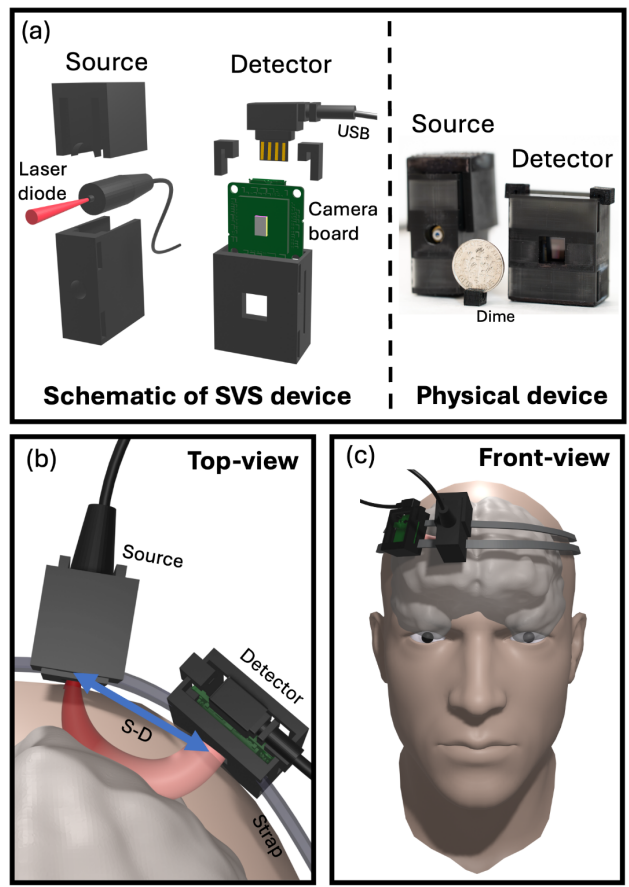

**Fig. 1** Compact speckle visibility spectroscopy (SVS) setup. (a) Design of the SVS device, consisting of a laser diode (source) and a CMOS-based board camera (detector) both housed in a 3D-printed mount. (left panel) 3D schematics breakdown. (right panel) Photograph of the actual SVS device. (b) Top-view and (c) front-view schematics illustrating

the SVS device in use on a subject's forehead. When set at a specific S-D distance, the SVS device can effectively measure cerebral blood flow.

The experimental SVS processing analysis flowchart for deriving the CBF from recorded camera images is shown in Fig. 2. The squared speckle contrast $K_{raw}^2(I)$ of a recorded camera image $I$, Fig. 2(a) is calculated as:

$$K_{raw}^2(I) = \frac{\sigma^2(\tilde{I})}{\mu^2(\tilde{I})}, \tag{1}$$

where in Eq. (1), $\tilde{I} = I - I_{offset}$, with $I$ the recorded camera image and $I_{offset}$ the camera offset. To experimentally measure the camera offset, we capture a series of images without any source illumination, and then calculate the mean offset image $I_{offset}$. The variance of $\tilde{I}$ is $\sigma^2(\tilde{I})$ and the mean is $\mu(\tilde{I})$. This calculation does not account for noises that contribute to the variance of the images. To account for these noises, we use an adjusted squared speckle contrast $K_{adjusted}^2$, which is commonly calculated as[17,30–32]:

$$K_{adjusted}^2 = K_{raw}^2 - K_{shot}^2 - K_{quant}^2 - K_{cam}^2 - K_{sp}^2, \tag{2}$$

with $K_{shot}^2$ accounting for variance contributions from the shot noise, $K_{quant}^2$ for the variance inherited from quantization, $K_{cam}^2$ for the variance contributions of the camera's readout noise and dark noise, and $K_{sp}^2$ for the spatial inhomogeneities. See Fig. 2(b) for examples of raw and noise speckle contrast measurements. For each of the image $\tilde{I}$, they can be calculated as the following[17,30–32]:

$$K_{shot}^2(I) = \left(\frac{\gamma}{\mu(\tilde{I})}\right), \tag{3a}$$

$$K_{quant}^2(I) = \left(\frac{1}{12\mu(\tilde{I})^2}\right), \tag{3b}$$

$$K_{cam}^2(I) = \left(\frac{\sigma_{cam}^2}{\mu(\tilde{I})^2}\right), \tag{3c}$$

$$K_{sp}^2(I) = \left(\frac{\sigma_{sp}^2}{\mu(\tilde{I})^2}\right). \tag{3d}$$

In Eq. (3a), $\gamma$ is the analog to digital conversion ratio associated to the camera, which depends on the gain setting and the conversion factor $CF$ of the camera, as $\gamma = \frac{gain}{CF}$. In our investigations, the gain was set within a range of 1 to 72, corresponding to a 0 to 37 dB setting. The gain was tuned depending on the signal intensity. In 8-bit mode, the Basler camera we used had a conversion factor of $CF = 40.7$. To reduce quantization noise, the gain of the camera was adjusted such that the camera readout grayscale values fell within the range of 40 to 255 at 8-bit recording unless the signal is too low. The camera noise $\sigma_{cam}^2$ was estimated by calculating the variance of a series of 500 camera images recorded in the absence of any illumination sources. The measured camera noise may introduce an offset bias, which is rectified by subtracting a bias term. The spatial variations in photon flux across the sensor area is accounted by $\sigma_{sp}^2$. With these calibrations, we can acquire the adjusted speckle contrast $K_{adjusted}^2$ for each of the recorded image.

After obtaining $K_{adjusted}^2$, one can calculate the decorrelation time $\tau$ as[15,17,33,34]:

$$K_{adjusted}^2 = \frac{\tilde{\beta}\tau}{T}\left[1 + \frac{\tau}{2T}\left(\exp\left(-\frac{2T}{\tau}\right) - 1\right)\right], \quad (4)$$

where $T$ is the exposure time of the camera, and $\tilde{\beta} = \beta - \beta_{offset}$ is a constant that accounts for the loss of correlation associated with the ratio of the detector size to the speckle size and polarization[33]. At low signal, $\beta$ may deviate from typical calibration due to high sensitivity to noise. To mitigate the issue that $K_{adjusted}^2 < 0$ in low signal situations, a correction term $\beta_{offset}$ enforces a positive speckle contrast. With our proposed compact SVS setup, we employ a lensless imaging configuration to enhance the numerical aperture, enabling the recording of multiple speckles within a single pixel, leading to $\beta \approx 0.05$, measured with a static sample. The measured $\beta$ value is relatively low because the average speckle size is smaller than the pixel size, resulting in multiple speckles per pixel. In SVS, the detecting device operates with an exposure time significantly greater than the decorrelation time of the sample, i.e. $T \gg \tau$. In our case, $T = 6$ ms. Consequently, Eq. (4) simplifies to:

$$K_{adjusted}^2 \approx \frac{(\beta - \beta_{offset})\tau}{T}. \quad (5)$$

The cerebral blood flow (CBF) can be related to $K_{adjusted}^2$ (and $\tau$) as [35,36]:

$$CBF = \frac{1}{K_{adjusted}^2} \approx \frac{T}{(\beta - \beta_{offset})\tau}. \quad (6)$$

See Fig. 2(c) for typical example of measured CBF dynamics with our SVS device. The CBF accounts for the total volume of blood moved in each time period. It can also be measured in blood flow index (BFI) metric[1].

The CBF metric accounts for the total volume of blood moved in a given time period. According to classical fluid mechanics and Poiseuille's law, blood flow can be expressed as $BF = \frac{\delta P \pi r^4}{8\eta L}$ where $\delta P$ represents the difference in blood pressure, $r$ denotes the radius of the blood vessel, $\eta$ is the dynamic viscosity of the blood, and L signifies the length of the blood vessel [1]. Thus, any alteration in the blood flow means that there is a change in either the blood pressure or a change in the diameter of the blood vessel. It is worth noting that even a slight adjustment in the blood vessels' radius can have a profound impact on blood flow due to the fourth power relationship with $r$. Such variations are especially significant as they accompany the modulation and regulation of CBF. In results shown later, we utilize the relative cerebral blood flow (rCBF) metric to provide normalized blood flow information for enhanced comparability across measurements.

Note that the computational requirements needed to calculate the speckle contrast in Eqs. (1)-(3) from the recorded camera images can be handled by a standard consumer-grade computer [e.g., AMD 7950X CPU]. The most resource-demanding step is the calculation of $K_{raw}^2$ in Eq. (1), as the noise terms in Eq. (3) only need to be calculated once (pre-calibration) or were already calculated in Eq. (1). Therefore, the data recorded from our SVS compact device can be processed and stored in real-time by using a dedicated Basler USB-PCIE card and SSD. It is also possible to expand the device to multiple channels.

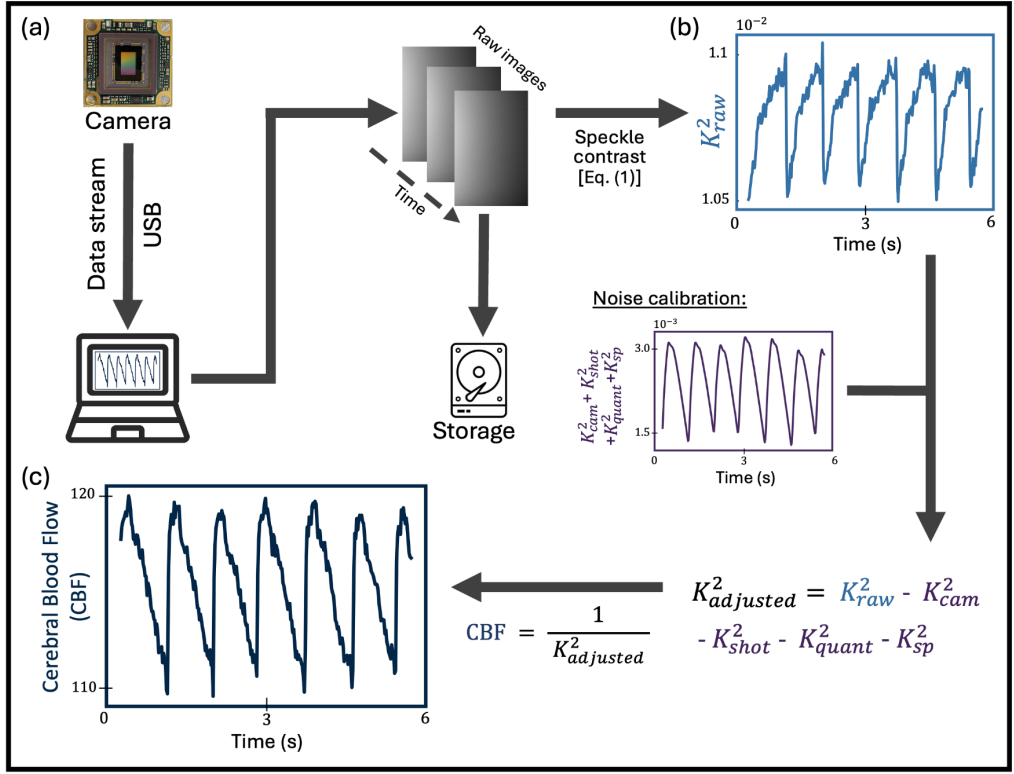

**Fig. 2** Compact SVS processing analysis flowchart for deriving the cerebral blood flow (CBF) from recorded camera images. (a) Recording and storing of SVS camera images. (b) Measured raw speckle contrast calculated from the images in (a). (c) Calculated CBF after calibrating the raw speckle contrast in (b).

.

## 3  Results and Discussion

Relative to the traditional fiber-based SVS systems[15–17], the compact SVS arrangement, where the sensor is directly positioned atop the region of interest offers the larger collection area and numerical aperture of the sensor allow for two orders of magnitude increase in the number of photons collected. To demonstrate the superior signal strength and stability of the compact SVS over traditional fiber-based SVS systems, we compared the two systems. The experimental configuration is shown in Fig. 3(a) and features a continuous-wave 785 nm laser diode, acting as a common light source, and two SVS detection modules symmetrically placed on each side of the laser source at the same S-D separation distance. On one detection side, the compact SVS was composed of a board camera [Basler daA1920-160um], directly positioned on the sample. On the opposing detection side, the fiber SVS was composed of a 600-um diameter multimode optical fiber [Thorlabs FT600UMT], positioned on the sample. The other end of the fiber was coupled onto an identical camera to the one used in the compact SVS[16].

Theoretically, we expect the compact version to yield a signal gain of about 70 times compared to the fiber, as a result of the increased collection area. The camera sensor's dimension is 6.6 mm by 4.1 mm, resulting in an approximate sensor area of 27 mm$^2$ compared to the 0.28 mm$^2$ area of the 600-um diameter multimode optical fiber, leading to about 95 times gain in detecting area. However, the camera sensor is positioned with a 5 to 7 mm gap from the sample, while the fiber is directly placed in contact with the sample. In this configuration, we calculated the numerical aperture of the camera to be $NA_x = 0.28$ on one dimension and $NA_y = 0.42$ on the other dimension. The fiber has a numerical aperture of $NA_{fiber} = 0.39$. By taking into account the NA difference between the camera and fiber, we expect the collected signal gain between the compact SVS over fiber SVS to be $\frac{NA_x \cdot NA_y}{(NA_{fiber})^2} \cdot \frac{Area_{CMOS}}{Area_{fiber}} \approx 75$ times.

To experimentally validate this gain, we used a static sample (a thick slice of meat) and measured the camera readout signal at different S-D distances for the two detection units. The S-D distances ranged from 1.5 cm (2.5 cm) for the fiber (compact) SVS to 10.5 cm. The mounting encasing units prevent smaller S-D distances. The results are presented in Fig. 3(b) and were averaged over six different realizations at different locations. The error bars were estimated by calculating the standard deviation over the six different realizations. As shown, a consistent gain in the number of photons is captured for the compact SVS over the fiber SVS system across the multiple S-D distances. By calculating the signal ratio of the two, we determined that the compact version capture about 70 times more signal than its fiber-based SVS counterpart. This significant improvement leads to an enhanced detectability at extended S-D distances, up to an S-D distance increase of 2.5 cm for the same signal readout in this case. Note that both devices reach the noise floor of the camera, although at different S-D distances. The fiber SVS reaches the noise floor level at an S-D distance of approximately 5.5 cm, whereas the compact SVS maintains a robust signal even at a S-D distance of 8.0 cm on a static sample. These results showcase the superiority of compact SVS over fiber SVS in the ability to collect more signal, enabling the detectability of CBF at extended S-D distances. Note that this increase has not yet consider the increase in stability by eliminating the potential fiber movement during the camera exposure time.

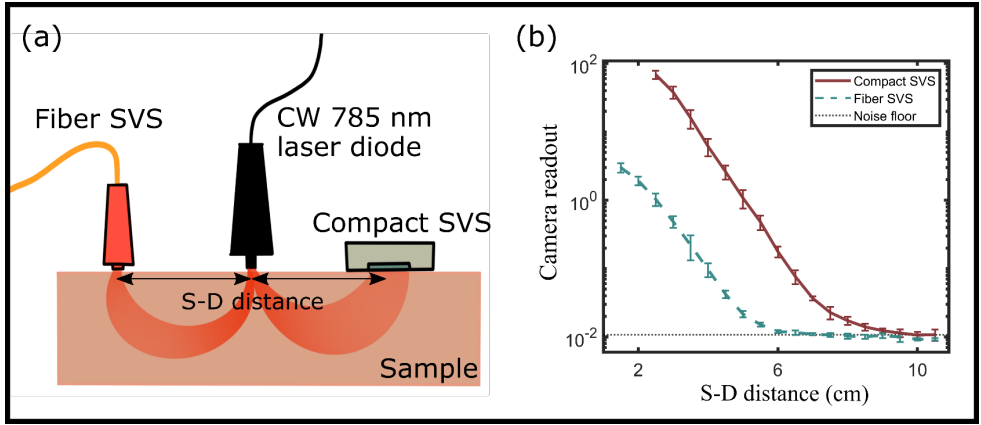

**Fig. 3** Experimental comparison between fiber SVS and compact SVS: (a) Overview of the experimental setup. (b) Camera readout signal intensity for fiber and compact SVS measured at various S-D distances on a static sample. Notably, the compact SVS demonstrates a significantly higher readout signal, averaging approximately 70 times more than its fiber counterpart. The fiber SVS reaches the noise floor level at an S-D distance of approximately 5.5 cm, whereas the compact SVS maintains a robust signal even at a S-D distance of 8.5 cm.

Next, we characterized the stability of the compact SVS over the fiber SVS. For that, we designed two distinct experiments. In the first experiment, presented in Fig. 4(a), both the compact and fiber SVS systems were affixed on top of an one-layer phantom, which comprised of a sealed container filled with a liquid mixture (3D printing resin)[16]. The liquid mixture was positioned on an orbital shaker set at a rotating speed of 90 rotations per minute[16]. Both compact SVS and fiber SVS systems rotated synchronously with the liquid mixture. During each rotation, the SVS systems are measuring the change of decorrelation time within the liquid mixture[16]. In this context, the SVS systems measure a liquid flow dynamic similar to the blood flow dynamic for humans, see Appendix A of Ref 16. The compact SVS and fiber SVS were set at S-D distances that yielded an equivalent photon count. In this configuration, the light power collected by the compact SVS and fiber SVS detection systems are equivalent.

Consequently, we anticipate assessing the stability of the compact SVS and fiber SVS by comparing the recorded liquid flow from the rotating phantom. The results are shown in Fig. 4(b). As expected, the liquid flow measured by the compact SVS exhibits superior signal quality compared to that obtained by the fiber SVS. This observation is further validated when examining the frequencies present in the Fourier spectrum of the flow signal intensity $I$, Fig. 4(c). The Fourier amplitude peak, centered around 1.5 Hz, corresponds to the rotational frequency of the orbital shaker (90 rotations per minute, translating to 1.5 rotations per second). The peak at around 3 Hz (4.5 Hz), representing the second (third) harmonic of the flow pulsation[37], is observable in both the compact SVS and fiber SVS spectra. However, the noise level is slightly higher in the fiber SVS Fourier spectrum, indicating that the measured flow intensity from the fiber SVS is less reproducible than that from compact SVS. To quantitatively evaluate the reproducibility of the measured liquid flow signal, we computed the Pearson correlation factor for each SVS system as:

$$\rho\big(I(t), I(t+dt)\big) = \frac{\sum_{t=1}^{T}(I(t)-\bar{I})(I(t+dt)-\bar{I})}{\sqrt{\sum_{t=1}^{T}(I(t)-\bar{I})^2 \sum_{t=1}^{T}(I(t+dt)-\bar{I})^2}}, \tag{5}$$

where $I(t)$ is the measured flow signal in Fig. 4(a), $\bar{I}$ is the mean flow intensity, and $I(t + dt)$ is the signal shifted by one period of $dt = 1/1.5\ Hz = 0.66$ sec. The compact SVS correlation factor was $\rho_{compact} = 0.94$ and the fiber SVS was $\rho_{fiber} = 0.67$, demonstrating that the measured periodic signal from the compact SVS is significantly more stable than its fiber counterpart. We further investigated the compact SVS's robustness against human head movements.

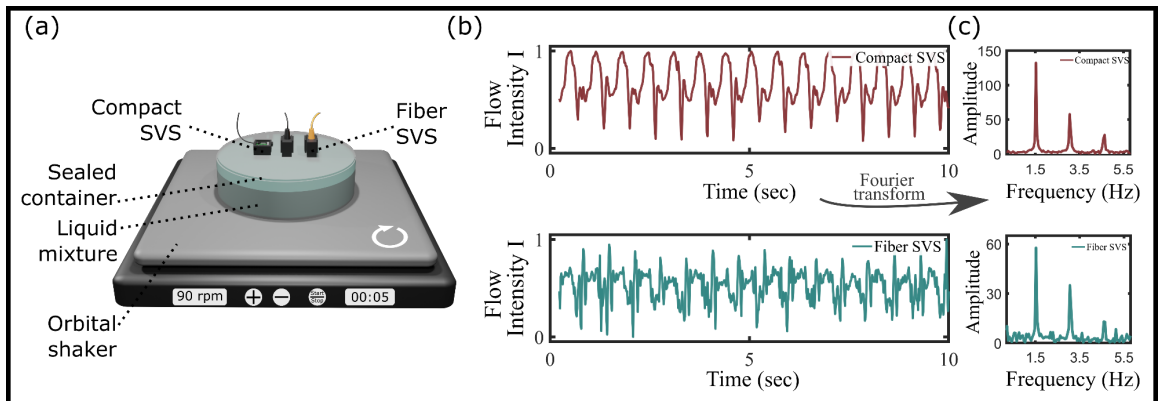

**Fig. 4** Experimental comparison of flow reproducibility between fiber SVS and compact SVS by using a one-layer phantom rotating on an orbital shaker. (a) Experimental arrangement. (b) Measured flow intensity. (c) Rotating frequencies obtained by Fourier transforming the flow intensity signal in (b).

The second experiment, presented in Fig. 5(a), entailed an evaluation of head movement instabilities. To conduct this assessment, both SVS systems were positioned on the forehead of a subject, with a static scattering block interposed between the SVS setups and the subject's forehead. The static scattering block comprised a rigid block of packaging foam, complemented by a thick layer of black tape on its backside to prevent any laser light from entering the subject's head. Consequently, the SVS systems exclusively detected light interacting with the static scattering block. The S-D distances of both the compact SVS and fiber SVS are equivalent, in order to replicate CBF data acquisition scenarios. As a result, the signal intensity on the fiber SVS is about 70 times lower than that of the compact SVS. The measurement was performed over a 30-second interval, following a specific protocol: from 0 to 10 seconds, the subject maintained a still position; at the 10-second mark, the subject was asked to laterally move their head (left to right and right to left) for the subsequent 10 seconds; and finally, the subject resumed a stationary position for the remaining 10 seconds. The results are presented in Fig. 5(b) and show that the flow measured by the compact SVS exhibits less noise movement that with the fiber SVS during head movements. In addition, the overall flow intensity notably rises due to the SVS systems' movements accompanying head motions. As shown in Fig. 5(b), this increase in flow intensity is more pronounced for the fiber SVS than the compact SVS, indicating than the compact SVS experiences less movement vibrations than the fiber SVS during head motions. The more prominent shift in the fiber SVS is due to additional decorrelation resulting from fiber movement, which creates blurrier speckle images. These blurrier images correspond to lower contrast, leading to higher flow intensity as explained in Eq. 5 and Eq. 6. By removing the fiber, this source of instability is largely eliminated, resulting in the compact SVS displaying more stable flow measurements during movements.

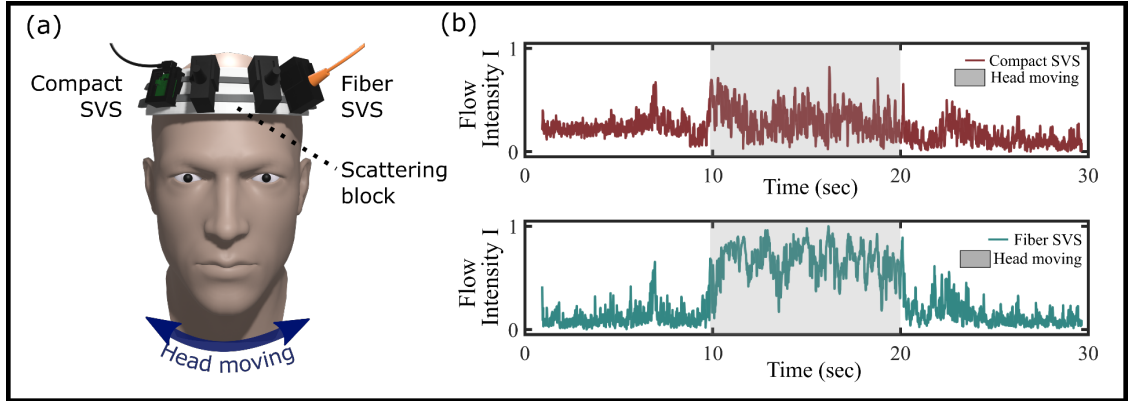

**Fig. 5** Experimental comparison of stability between fiber SVS and compact SVS by using a static scattering block on the forehead of a subject. (a) Experimental arrangement. (b) Measured flow intensity when the head is moving from 10 to 20 sec.

To determine the capability of the compact SVS system to detect blood flow at large S-D distances, we further tested the two SVS systems by measuring CBF on the forehead of a human subject at S-D distances ranging from 3 cm to 5.5 cm. The results are shown in Fig. 6. Figure 6(a) shows the measured CBF by the compact and fiber SVS system at S-D distances of 3 cm, 4 cm, and 5 cm. Figure 6(b) shows the normalized Fourier transform of the blood flow signal in Fig. 6(a). The Fourier amplitude peak centered around HR = 1 Hz corresponds to the heart rate amplitude peak of the subject[16]. We verified that the heart rate of the subject matches with the one measured from a standard pulse oximeter.

The quality of the measured blood flow signal can be assessed by examining the amplitude of the heart rate Fourier peak[16]. Figure 6(c) shows the amplitude of the heart rate Fourier peak at different S-D distances ranging from 3 cm to 5.5 cm for both the compact and fiber SVS. The results were averaged over three realizations. As shown, the compact SVS exhibits a significant gain over the fiber SVS system across the S-D distances. Finally, the heartbeat of the subject can be measured by measuring the frequency of the heart rate $freq_{HR}$ Fourier peak[16]. The measured SVS heart rate $freq_{HR}^{SVS}$ can be compared with the one measured from a standard pulse oximeter $freq_{HR}^{oxymeter}$ by calculating the relative percentage error as:

$$\Delta freq_{HR} = \frac{\left|freq_{HR}^{oxymeter} - freq_{HR}^{SVS}\right|}{freq_{HR}^{oxymeter}} \cdot 100\% \tag{6}$$

Figure 6(d) shows the heart rate relative percentage error across the S-D distances for both the compact and fiber SVS systems. As shown, the compact SVS exhibits a lower error than the fiber SVS system within S-D distances from 3 cm to 5 cm.

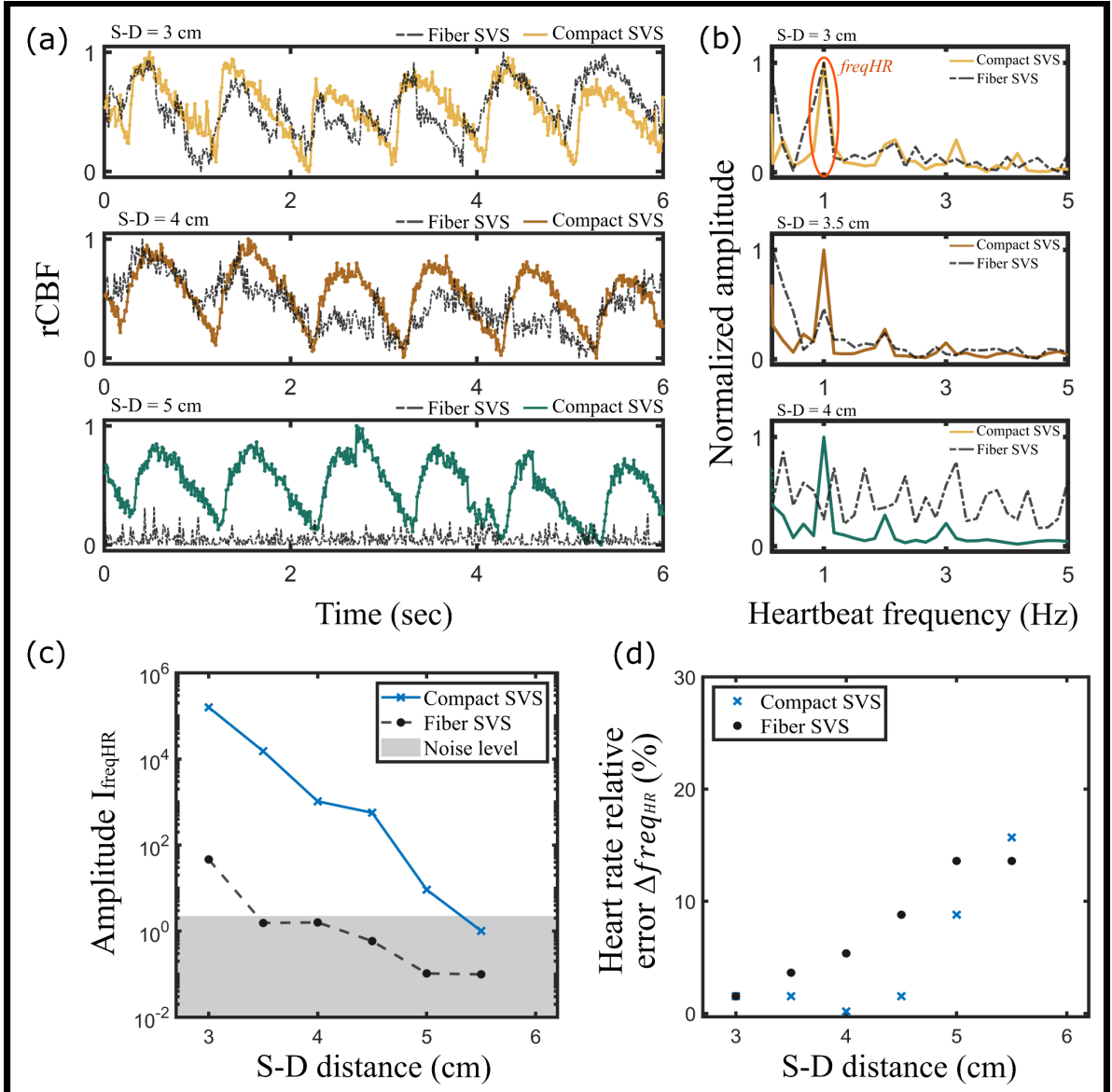

**Fig. 6** Experimental comparison between fiber SVS and compact SVS in CBF measurement. (a) Measured CBF at different S-D distances on the forehead of a subject. (b) Heart rate frequencies obtained by Fourier transforming the CBF signal in (a). (c) Amplitude of the heartbeat frequency peak $I_{freqHR}$ as a function of the S-D distance. (d) Heart rate relative percentage error between SVS and with a pulse oximeter. In (c) and (d), the results were averaged over three realizations.

Finally, the measurements of Fig. 6 were repeated on a cohort of five subjects. The results are shown in Fig. 7, by showing the amplitude of the heart rate Fourier peak and the heart rate relative error at different S-D distances for both the compact and fiber SVS. For each subject, three realizations were recorded. The results were averaged over the five different subjects and the error bar were determined by calculating the standard deviation. Across the five subjects, compact SVS shows a significant gain in signal over the fiber SVS, measuring CBF up to a S-D distance of 5 cm. Based on Fig. 7(a), the amplitude $I_{freqHR}$ for compact SVS at S-D distance of 5 cm is comparable to 3 cm from fiber SVS. This is further validated by the correct prediction of the heart rate frequency at large S-D distances up to 5 cm, as shown in Fig. 7(b). The pronounced error bars observed at larger S-D distances result from the signal being heavily masked by noise, leading to inconsistent and unreliable heart rate predictions across different realizations and subjects. The typical measured CBF signals across the five subjects are shown in Appendix figure Fig. B1.

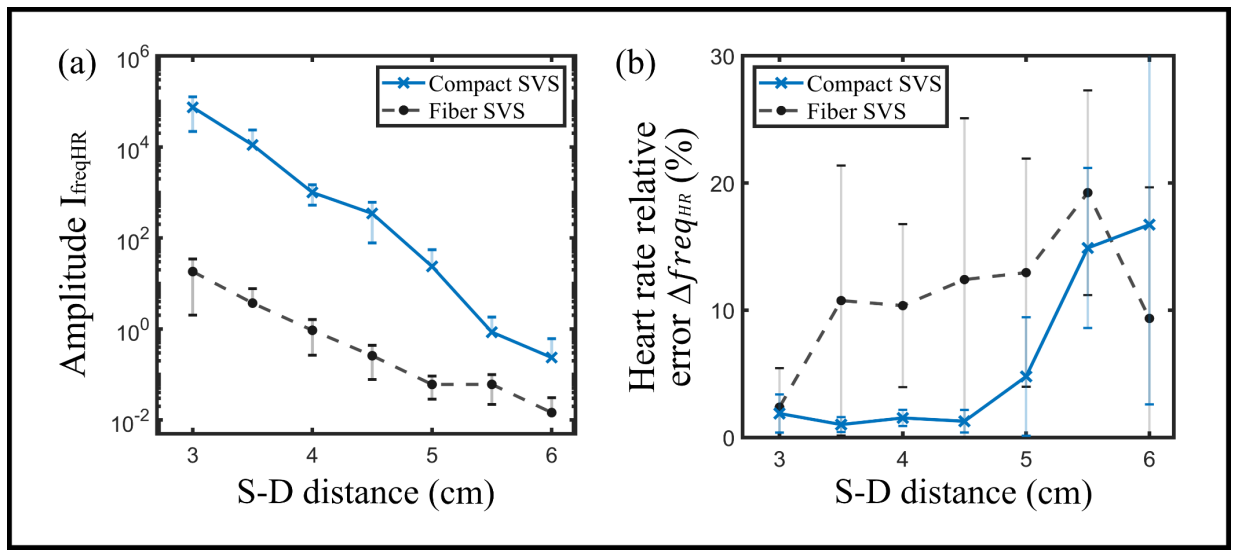

**Fig. 7** Experimental comparison between fiber and compact SVS in CBF measurement on a cohort of five subjects. (a) Averaged amplitude of the heartbeat frequency peak and (b) heart rate relative error as a function of the S-D distance.

## 4  Conclusion

To summarize, we presented a compact and cost-effective laser-powered device for assessing cerebral blood flow (CBF). The device is based on speckle visibility spectroscopy (SVS) technology, which is an off shoot of laser speckle contrast imaging (LSCI). The hardware consisted solely of only two components: a carefully chosen laser diode and a meticulously selected CMOS-based board camera. It offers real-time CBF monitoring at 80 Hz sampling rate while maintaining a lightweight and modular design, and achieving 70 times increase in collected signal over typical fiber-based SVS models. We demonstrated that the device could measure cerebral blood flow up to a source-to-detector distance of 4.5 cm across a cohort of five subjects. For future investigations, we plan on showcasing the capability of our device to assess cerebrovascular reactivity, by measuring the ability of the brain to adjust CBF in response to oxygen supply changes within the body. By characterizing these dynamic brain responses, including the extent of changes in CBF and the speed at which the brain returns to a baseline level of activity, we aim to assess the cerebrovascular health of a participant and evaluate their risk of experiencing a cerebrovascular disease.

**Appendix A: Study participants**

We tested and experimented our device across a cohort of five subjects. Participants for this study were recruited from the Caltech and Pasadena community, selected among adult humans aged from 21 to 65 years. Prior to the experiments, each participant completed a health questionnaire, and their blood pressure was recorded. The human research protocol for this study received approval from the Caltech Institutional Review Board (IRB).

To simplify the experiment and implementation of the device, SVS was conducted on hairless areas, such as the forehead or frontotemporal region. Optical transmission is optimal when both the light source and detector are positioned on hairless regions of the head. An ideal scenario entails a hair-free circular space of 0.5 cm diameter for the illumination spot and a square area of 1 cm × 1 cm for the detection device. While these requirements are manageable, they may pose inconveniences for participants with hair who are unwilling to shave. To mitigate this challenge, as further step, we aim to design and use of 3D-printed mounts equipped with hair separators. This innovative solution aims to minimize hair interference with the optical transmission process, enhancing the device's usability and accommodating participants with hair.

**Appendix B: CBF measurement results on the subjects**

In this section, we show in Fig. B1 an example of the CBF measurement results from the compact SVS system used on the five subjects of Fig. 7.

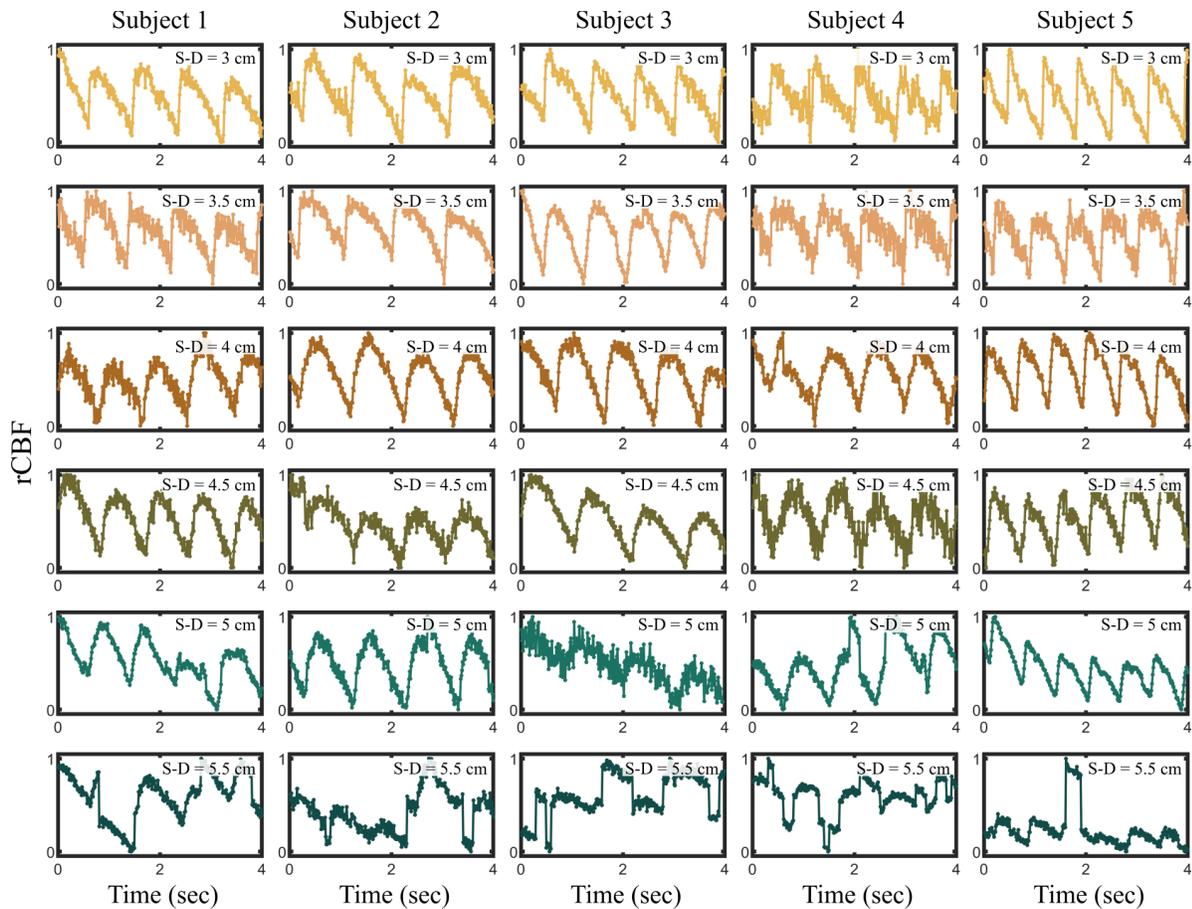

**Fig. B1** CBF measurement results on the forehead of the five subjects in Fig. 7.

## Acknowledgments

The authors thank Professor Jerome Mertz, Dr. Kate Bechtel, Dr. Cody Dunn, and the Rockley Photonics team for helpful discussions. The authors thank Ruizhi Cao for his help in designing the schematic figure. This research was supported by the National Institutes of Health — Award No. 5R21EY033086-02. Simon Mahler is the recipient of the 2024 SPIE-Franz Hillenkamp Postdoctoral Fellowship.